\documentclass[axioms,article,accept,pdftex,moreauthors]{Definitions/mdpi} 

\firstpage{1} 
\pubvolume{13}
\issuenum{3}
\articlenumber{160}
\pubyear{2024}
\copyrightyear{2024}
\externaleditor{Academic Editor: Mariam Zomorodi}
\datereceived{25 January 2024 } 
\daterevised{25 February 2024 } 
\dateaccepted{25 February 2024 } 
\datepublished{29 February 2024} 
\hreflink{https://doi.org/10.3390/\\axioms13030160} 

\usepackage{amsfonts}    
\usepackage{amsmath}
\usepackage{amsthm,amssymb}
\usepackage{threeparttable} 
\usepackage[markup=underlined]{changes}

\newcommand{\vect}[1]{\boldsymbol{#1}} 

\usepackage{booktabs} 
\usepackage{mathtools}
\usepackage{mathrsfs}

\newcommand{\beq}{\begin{equation}}
\newcommand{\eeq}{\end{equation}}
\newcommand{\ga}{\lower.7ex\hbox{$\;\stackrel{\textstyle>}{\sim}\;$}}
\newcommand{\la}{\lower.7ex\hbox{$\;\stackrel{\textstyle<}{\sim}\;$}}

\Title{A Comparison between Invariant and Equivariant Classical and Quantum Graph Neural Networks}

\TitleCitation{A Comparison between Invariant and Equivariant Classical and Quantum Graph Neural Networks}

\Author{
{Roy T.~Forestano} 
 $^{1,}$*\orcidA{}, 
Marçal Comajoan Cara $^{2}$\orcidB{},
Gopal Ramesh Dahale $^{3}$\orcidC{},
Zhongtian Dong $^{4}$\orcidG{},\linebreak
Sergei Gleyzer $^{5}$\orcidH{},
Daniel Justice $^{6}$\orcidI{},
Kyoungchul Kong $^{4}$\orcidJ{},
Tom Magorsch $^{7}$\orcidK{},
{Konstantin T.~Matchev} 
 $^{1}$\orcidD{}, 
Katia~Matcheva $^{1}$\orcidE{} 
and 
Eyup B.~Unlu $^{1}$\orcidF{} }

\AuthorNames{Roy T. Forestano, Marçal Comajoan Cara, Gopal Ramesh Dahale, Zhongtian Dong, Sergei Gleyzer, Daniel Justice, Kyoungchul Kong, Tom Magorsch, Konstantin T. Matchev, Katia Matcheva and Eyup B. Unlu}

\AuthorCitation{{Forestano, R.T.;} 
 Comajoan Cara, M.; Dahale, G.R.; Dong, Z.; Gleyzer, S.; Justice, D.; Kong, K.; Magorsch, T.; Matchev, K.T.; Matcheva, K.; et al.}

\address{$^{1}$ \quad {Institute} 
 for Fundamental Theory, Physics {Department}, University of Florida, Gainesville, FL 32611, USA; matchev@ufl.edu (K.T.M.); matcheva@ufl.edu (K.M.); eyup.unlu@ufl.edu (E.B.U.)\\
$^{2}$ \quad {Department}
~of Signal Theory and Communications, Polytechnic University of Catalonia, 08034~Barcelona,~Spain; marcal.comajoan@estudiantat.upc.edu \\
$^{3}$ \quad Indian {Institute} of Technology Bhilai, Kutelabhata, Khapri, District{-}
Durg, Chhattisgarh 491001, India; gopald@iitbhilai.ac.in \\
$^{4}$ \quad Department of Physics \& Astronomy, University of Kansas, Lawrence, KS 66045, USA; cdong@ku.edu (Z.D.); kckong@ku.edu (K.K.)\\
$^{5}$ \quad Department of Physics \& Astronomy, University of Alabama, Tuscaloosa, AL 35487, USA; sgleyzer@ua.edu \\
$^{6}$ \quad Software Engineering Institute, Carnegie Mellon University, 4500 Fifth Avenue, Pittsburgh, PA 15213, USA; dljustice@sei.cmu.edu \\
$^{7}$ \quad Physik-Department, Technische {Universit\"at }M\"unchen, James-Franck-Str. 1, 85748 Garching, Germany; tom.magorsch@tum.de
}

\corres{Correspondence: roy.forestano@ufl.edu}

\abstract{Machine learning algorithms are heavily relied on to understand the vast amounts of data from high-energy particle collisions at the CERN Large Hadron Collider (LHC). The data from such collision events can naturally be represented with graph structures. Therefore, deep geometric methods, such as graph neural networks (GNNs), have been leveraged for various data analysis tasks in high-energy physics. One typical task is jet tagging, where jets are viewed as point clouds with distinct features and edge connections between their constituent particles. The increasing size and complexity of the LHC particle datasets, as well as the computational models used for their analysis, have greatly motivated the development of alternative fast and efficient computational paradigms such as quantum computation. In addition, to enhance the validity and robustness of deep networks, we can leverage the fundamental symmetries present in the data through the use of invariant inputs and equivariant layers. In this paper, we provide a fair and comprehensive comparison of classical graph neural networks (GNNs) and equivariant graph neural networks (EGNNs) and their quantum counterparts: quantum graph neural networks (QGNNs) and equivariant quantum graph neural networks (EQGNN). The four architectures were benchmarked on a binary classification task to classify the parton-level particle initiating the jet. Based on their
{area under the curve (AUC)}
 scores, the quantum networks were found to outperform the classical networks. However, seeing the computational advantage of quantum networks in practice may have to wait for the further development of quantum technology and its associated 
 {application programming interfaces (APIs)}
.}


\begin{document}

\section{Introduction}

Through the measurement of the byproducts of particle collisions, the~Large Hadron Collider (LHC) collects a substantial amount of information about fundamental particles and their interactions. The~data produced from these collisions can be analyzed using various supervised and unsupervised machine learning methods~\cite{Andreassen2019,Shlomi2021,Mikuni2021,mokhtar2022graph,Mikuni2020}. Jet tagging is a key task in high-energy physics, which seeks to identify the likely parton-level particle from which the jet originated.
By viewing individual jets as point clouds with distinct features and edge connections between their constituent particles, a~graph neural network (GNN) is considered a well-suited architecture for jet tagging~\cite{Mikuni2021,Shlomi2021}.

Classified as deep geometric networks, GNNs have the potential to draw inferences about a graph structure, including the interactions among the elements in the graph~\cite{velickovic2023connected,ZHOU202057}. Graph neural networks are typically thought of as generalizations of convolutional neural networks (CNNs), which are predominantly used for image recognition, pattern recognition, and~computer vision~\cite{kipf2016semi,kipf2017semisupervised}. This can be attributed to the fact that in~an image, each pixel is connected to its nearest neighboring pixels, whereas in a general dataset, one would ideally like to construct an arbitrary graph structure among the samples. Many instances in nature can be described well in terms of graphs, including molecules, maps, social networks, and~the brain. For~example, in~molecules, the~nodal data can be attributed to the atoms, the~edges can be characterized as the strength of the bond between atoms, and~the features embedded within each node can be the atom's characteristics, such as~reactivity. 

Generally, graphically structured problems involve unordered sets of elements with 
{a} learnable embedding of the input features. Useful information can be extracted from such
 {graphically}
 structured data by 
 embedding them within GNNs. Many subsequent developments have been made to GNNs since their first implementation in 2005
 {. These developments have included}
 graph convolutional, recurrent, message passing, graph attention, and~graph transformer architectures~\cite{velickovic2023connected,Shlomi2021,gilmer2017neural,velickovic2018graph}. 
  
To enhance the validity and robustness of deep networks, invariant and equivariant networks have been constructed to learn the symmetries embedded within a dataset by preserving an oracle in the former and by enforcing weight sharing across filter orientations in the latter~\cite{lim2022equivariant,ecker2018a}. Utilizing analytical invariant quantities characteristic of physical symmetry representations, computational methods have successfully rediscovered fundamental Lie group structures, such as the $SO(n)$, $SO(1,3)$, and~$U(n)$ groups~\cite{Forestano:2023fpj,Forestano:2023qcy,Forestano:2023ijh,Forestano:2023edq}. Nonlinear symmetry discovery methods have also been applied to classification tasks in data domains~\cite{Roman:2023ypv}. 
The simplest and most useful embedded symmetry transformations include translations, rotations, and~reflections, which have been the primary focus in invariant (IGNN) and equivariant (EGNN) graph neural networks \cite{maron2019invariant,Gong:2022lye,satorras2022en}. 

  
The learned representations from the collection of these network components can be used to understand unobservable causal factors, uncover fundamental physical principles governing these processes, and~possibly even discover statistically significant hidden anomalies. 
However, with~increasing amounts of available data and the computational cost of these deep learning networks, large computing resources will be required to efficiently run these machine learning algorithms. The~extension of classical networks, which rely on bit-wise computation, to~quantum networks, which rely on qubit-wise computation, is already underway as a solution to this complexity problem. 
{Due to superposition and entanglement among qubits, quantum networks are able to store the equivalent of $2^n$ characteristics from $n$ two-dimensional complex vectors. In~other words, while the expressivity of the classical network scales linearly, that of the quantum network scales exponentially with the sample size $n$ \cite{Preskill:2018jim}.} 
Many APIs, including Xanadu's Pennylane, Google's Cirq, and~IBM's Qiskit, have been developed to allow for the testing of the quantum circuits and quantum machine learning algorithms running on these quantum~devices.

In the quantum graph structure, classical nodes can be mapped to 
{the quantum states of the qubits}
, real-valued features to the complex-valued 
{entries}
 of the states, edges to the interactions between states, and~edge attributes to the strength of the interactions between the quantum states. Through a well-defined Hamiltonian operator, the~larger structure of a classical model can then be embedded into the quantum model. 
 {The unitary operator constructed from this}
 parameterized Hamiltonian determines the 
 {temporal} evolution of the quantum system by acting on the 
 {fully entangled} quantum 
 {state of}
  the graph. Following several layers of application, a~final state measurement of the quantum system can then be made to reach a final prediction. The~theory and application of unsupervised and supervised learning tasks involving quantum graph neural networks (QGNNs), quantum graph recurrent neural networks (QGRNNs), and~quantum graph convolutional neural networks (QGCNNs) have already been developed~\cite{PhysRevA.108.012410,verdon2019quantum}. Improvements to these models to arbitrarily sized graphs have been made with the implementation of ego-graph-based quantum graph neural networks (egoQGNNs) \cite{ai2023decompositional}. Quantum analogs of other advanced classical architectures, including generative adversarial networks (GANs), transformers, natural language processors (NLPs), and~equivariant networks, have also been proposed~\cite{PhysRevA.108.012410,niu2021entangling,10096772,disipio2021dawn, cherrat2022quantum,Meyer2023,nguyen2022theory,schatzki2022theoretical}. 

With the rapid development of quantum deep learning, this paper intends to offer a fair and comprehensive comparison between 
{classical} GNNs and their quantum counterparts. To~classify whether a particle jet has originated from a quark or a gluon, a~binary classification task was carried out using four different architectures. These architectures included a GNN, SE(2) EGNN, QGNN, and~permutation EQGNN. Each quantum model was fine tuned to have an analogous structure to its classical form. In~order to provide a fair comparison, all models used similar hyperparameters as well as a similar number of total trainable parameters. The~final results across each architecture 
{were recorded using identical training, validation, and~testing sets.}
 We found that QGNN and EQGNN outperformed their classical analogs on the particular binary classification task described above. Although~these results seem promising for the future of quantum computing, 
{the} further development of quantum 
{APIs}
 is required to allow for more general implementations of quantum architectures. 

\section{Data}

\begin{figure}[H]
\begin{adjustwidth}{-\extralength}{0cm}
  \centering
  \includegraphics[width=0.42\textwidth]{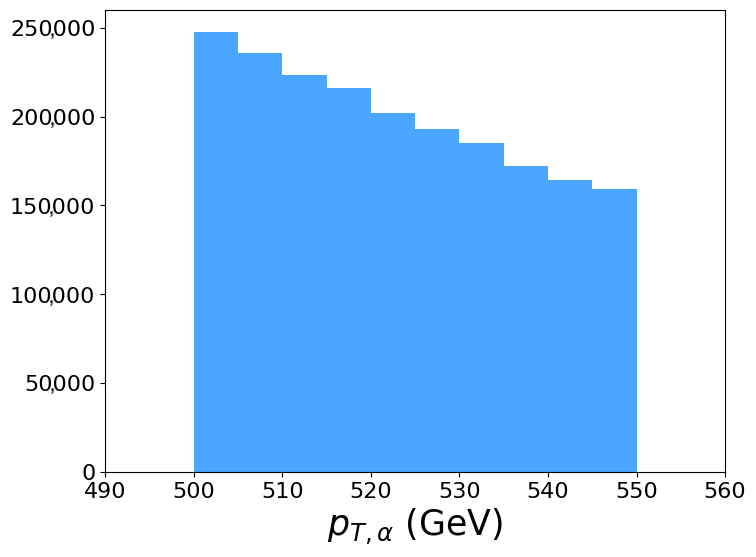}
  \includegraphics[width=0.42\textwidth]{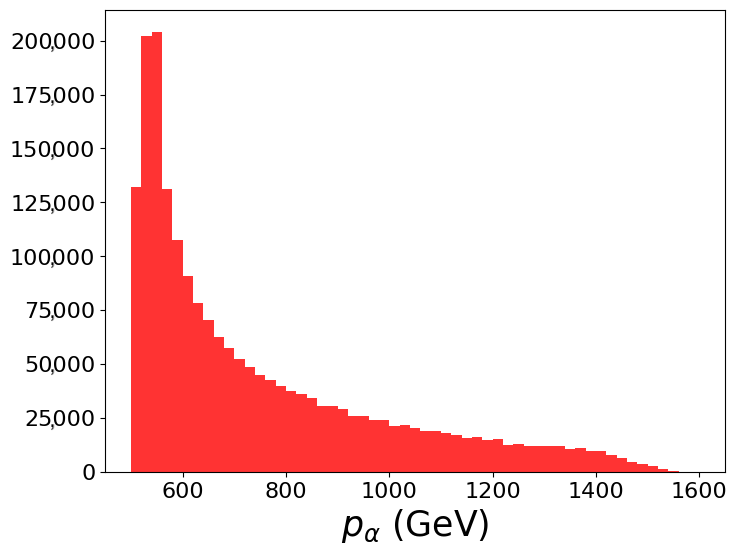}
  \includegraphics[width=0.42\textwidth]{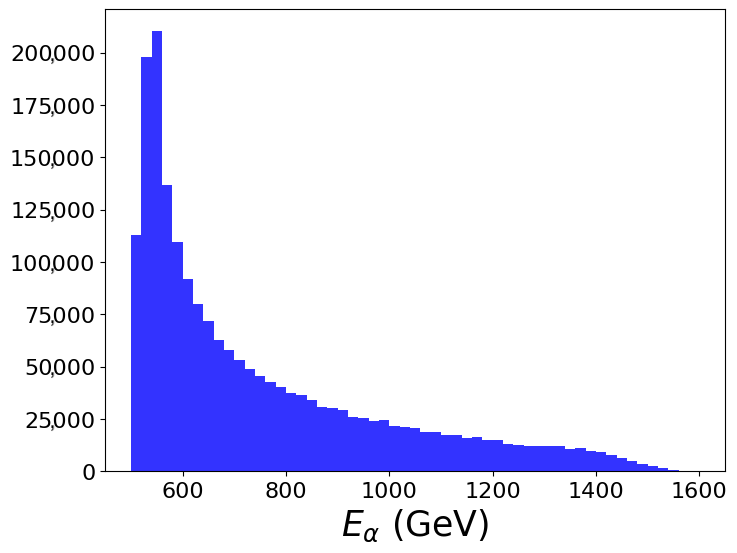}
\end{adjustwidth}
  \caption{{Distributions} 
 of the jet transverse momenta $p_T$, total momenta $p$, and~energies $E$.}
  \label{energy_momentum_distributions}
\end{figure}

The jet tagging binary classification task is illustrated with the high-energy physics (HEP) dataset \textit{Pythia8 Quark and Gluon Jets for Energy Flow} \cite{Komiske2019}
{. This dataset contains data from two million particle collision jets split equally into one million jets that originated from a quark and one million jets that originated from a gluon.}
 These jets resulted from LHC collisions with total center of mass energy $\sqrt{s} = 14$ TeV and were selected to have transverse momenta $p_T^{jet}$ between $500$ to $550$ GeV and rapidities $|y^{jet}| < 1.7$. The~jet kinematic distributions are shown in Figure~\ref{energy_momentum_distributions}. For~each jet $\alpha$, the~classification label is provided as either a quark with $y_\alpha =1$ or a gluon with $y_\alpha =0$
 {. Each particle}
 $i$ within the jet is listed with its transverse momentum $p^{(i)}_{T,\alpha}$, rapidity $y_\alpha^{(i)}$, azimuthal angle $\phi_\alpha^{(i)}$, and~PDG id~$I_\alpha^{(i)}$. 

\subsection{Graphically Structured~Data}

\begin{figure}[H]
    \includegraphics[width=0.48\textwidth]{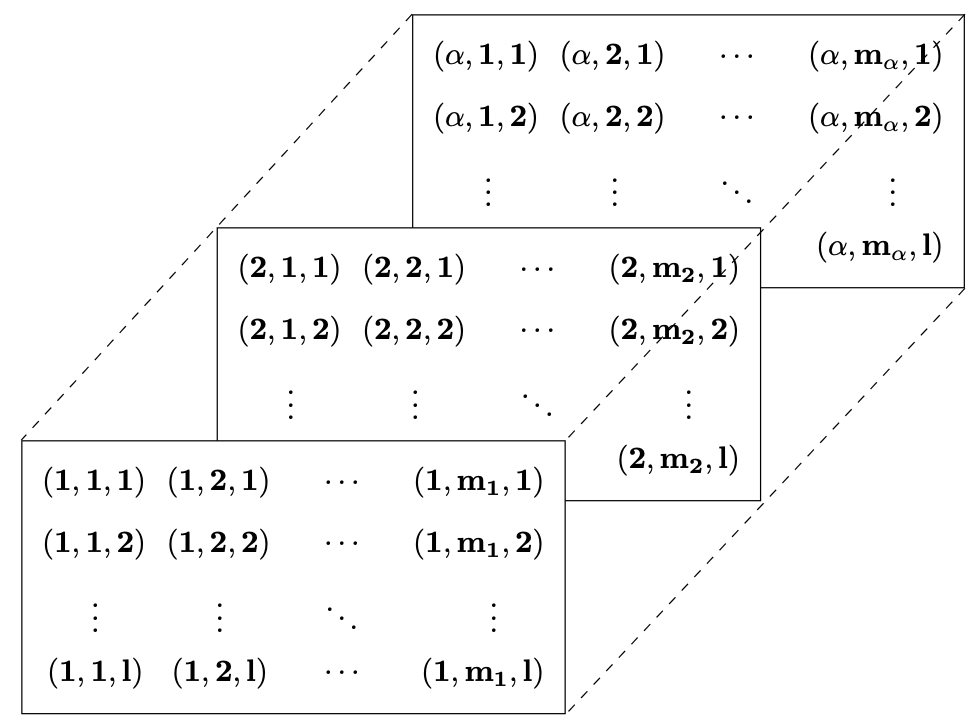}
    \includegraphics[width=0.48\textwidth]{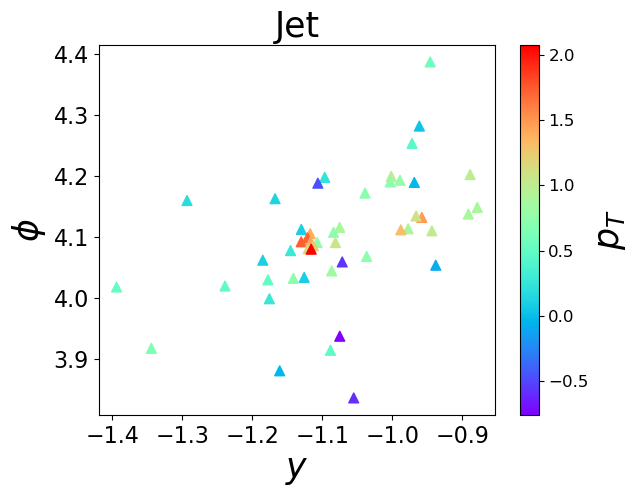}
    \caption{A visualization of graphically structured data (\textbf{left}) and a sample jet shown in the $(\phi,y)$ plane (\textbf{right}) with each particle color-coded by its transverse momentum $p_{T,\alpha}^{(i)}$.}
    \label{dataset_and_jet}
\end{figure}

A graph $\mathcal{G}$ is typically defined as a set of nodes $\mathcal{V}$ and edges $\mathcal{E}$, i.e.,~$\mathcal{G} = \{ \mathcal{V}, \mathcal{E} \}$. Each node $v^{(i)} \in \mathcal{V}$ 
is connected to its neighboring nodes $v^{(j)}$ via edges
$e^{(ij)}\in \mathcal{E}$. In~our case, each jet $\alpha$ can be considered as a graph $\mathcal{J}_\alpha$ composed of the jet's constituent particles as 
{the} nodes $v_\alpha^{(i)}$ with node features 
{$h_\alpha^{(il)}$}
 and edge connections $e_\alpha^{(ij)}$ between the nodes in $\mathcal{J}_\alpha$ with edge features $a_\alpha^{(ij)}$. 
It should be noted that the number of nodes within a graph can vary. This is especially true for the case of particle jets, where the number of particles within each jet can vary greatly. Each jet $\mathcal{J}_\alpha$ can be considered as a collection of $m_\alpha$ particles with $l$ distinct features per particle. An~illustration of graphically structured data and an example jet in the $(\phi,y)$ plane are shown in Figure~\ref{dataset_and_jet}.
\vspace{-6pt}

\subsection{Feature~Engineering}

We used the \texttt{Particle} package~\cite{eduardo_rodrigues_2023_8112280} to find the particle masses $m_\alpha^{(i)}$ from the respective particle IDs $I_\alpha^{(i)}$. 
From the available kinematic information for each particle $i$, we constructed new physically meaningful kinematic variables~\cite{Franceschini:2022vck}, which were used as additional features in the analysis. In~particular, we considered the transverse mass $m_{T,\alpha}^{(i)}$, the~energy $E_{\alpha}^{(i)}$, and~the Cartesian momentum components, $p_{x,\alpha}^{(i)}$, 
$p_{y,\alpha}^{(i)}$, and~$p_{z,\alpha}^{(i)}$, defined, respectively, {as}
\begin{align}
m_{T,\alpha}^{(i)} &= \sqrt{m_{\alpha}^{(i)2} + p_{T,\alpha}^{(i)2} }, \qquad 
E_{\alpha}^{(i)} = m_{T,\alpha}^{(i)} \text{cosh} ( y_{\alpha}^{(i)}), \label{eq:5features} \\
p_{x,\alpha}^{(i)} &= p_{T,\alpha}^{(i)} \text{cos} (\phi_\alpha^{(i)}), \qquad
p_{y,\alpha}^{(i)} = p_{T,\alpha}^{(i)} \text{sin} (\phi_\alpha^{(i)}), \qquad 
p_{z,\alpha}^{(i)} = m_{T,ij} \text{sinh} (y_\alpha^{(i)}).  \notag
\end{align}
{The} 
 original kinematic information in the dataset was then combined with the additional kinematic variables (\ref{eq:5features}) into a feature set $h_{\alpha}^{(il)}$, $l=0,1,2,\ldots,7$, as~follows:
\begin{equation}
h_{\alpha}^{(il)}
\equiv
\left\{
p^{(i)}_{T,\alpha}, 
y_\alpha^{(i)}, 
\phi_\alpha^{(i)},
m_{T,\alpha}^{(i)}, 
E_{\alpha}^{(i)}, 
p_{x,\alpha}^{(i)}, 
p_{y,\alpha}^{(i)}, 
p_{z,\alpha}^{(i)} 
\right\}.
\end{equation}
{These} features were then max-scaled by their maximum value across all jets $\alpha$ and particles $i$, i.e.,~$h_\alpha^{(il)} \to h_\alpha^{(il)}/max_{\alpha,i}(h_\alpha^{(il)})$.

Edge connections are formed via the Euclidean distance $\Delta R = \sqrt{\Delta \phi^2 + \Delta y^2}$ between one particle and its neighbor in $(\phi,y)$ space. Therefore, the~edge attribute matrix for each jet can be expressed as
\begin{align}
  a_\alpha^{(ij)} \equiv \Delta R_\alpha^{(ij)}= \sqrt{ \left( \phi_{\alpha}^{(i)} - \phi_{\alpha}^{(j)} \right)^2 + \left( y_\alpha^{(i)} - y_\alpha^{(j)} \right)^2 }.
  \label{eq:aij}
\end{align}

\subsection{Training, Validation, and~Testing~Sets}
We considered jets with at least $10$ particles. This left us with $N$ = 1,997,445 jets, 997,805 of which were quark jets. While the classical GNN is more flexible in terms of its hidden features, the~size of the quantum state and the Hamiltonian scale as $2^n$, where $n$ is the number of qubits. As~we shall see, the~number of qubits is given by the number of nodes $n_\alpha$ in the graph, i.e.,~the number of particles in the jet. Therefore, jets with large particle multiplicity require prohibitively complex quantum networks. 
{Thus, we limited ourselves to the case of $n_\alpha = 3$ particles per jet by only considering the three highest momenta $p_T$ particles within each jet.}
 In other words, each graph contained the set $\mathbf{h}_\alpha = (\mathbf{h}_\alpha^{(1)},\mathbf{h}_\alpha^{(2)},\mathbf{h}_\alpha^{(3)})$, where each $\mathbf{h}_\alpha^{(i)} \in \mathbb{R}^8$ and $\mathbf{h}_\alpha \in \mathbb{R}^{3 \times 8}$. For~training, we randomly picked $N$ = 12,500 jets and used the first 10,000 for training, the~next 1250 for validation, and~the last 1250 for testing. These sets happened to contain $4982$, $658$, and~$583$ quark jets, respectively.

\section{Models}

\textls[-25]{The four different models described below, including a GNN, an EGNN, a QGNN, and~an EQGNN,} were constructed to perform graph classification. The~binary classification task was determining whether a jet $\mathcal{J}_\alpha$ originated from a quark or a~gluon.

\subsection{Invariance and~Equivariance}

By making a network invariant or equivariant to particular symmetries within a dataset, a~more robust architecture can be developed. In~order to introduce invariance and equivariance, one must assume or learn a certain symmetry group $G$ of transformations on the dataset. A~function $\varphi : X \to Y$ is equivariant with respect to a set of group transformations $T_g : X \to X$, $g \in G$, acting on the input vector space $X$, if~there exists a set of transformations $S_g : Y \to Y$ that similarly transform the output space $Y$, i.e.,
\begin{align}
  \varphi(T_gx) = S_g \varphi(x).
\end{align}
{A} model is said to be invariant when, for all $g \in G$, $S_g$ becomes the set containing only the trivial mapping, i.e.,~$S_g = \{ \mathbb{I}_G \}$, where $\mathbb{I}_G \in G$ is the identity element of the group $G$~\cite{lim2022equivariant,esteves2020theoretical}. 

Invariance performs better as an input embedding, whereas equivariance can be more easily incorporated into the model layers. For~each model, the~invariant component corresponds to the translational and rotational invariant embedding distance $\varphi \equiv \Delta R_{\alpha}^{(ij)}$
{. Here, the~function $\varphi : \mathbb{R}^2 \times \mathbb{R}^2 \to \mathbb{R}$ makes up the edge attribute matrix $a_\alpha^{(ij)}$, as~defined in Equation~($\ref{eq:aij}$).}
 This distance is used as opposed to solely incorporating the raw coordinates. Equivariance has been accomplished through the use of simple nontrivial functions along with higher-order methods involving the use of spherical harmonics to embed the equivariance within the network~\cite{murnane2023equivariant,8100241}. Equivariance takes different forms in each model presented~here.

\subsection{Graph Neural~Network}
Classical GNNs take in a collection of graphs $\{ \mathcal{G}_\alpha \}$, each with nodes $v_\alpha^{(i)} \in \mathcal{V}_\alpha$ and edges $e_\alpha^{(ij)} \in \mathcal{E}_\alpha$, where 
 {each graph $\mathcal{G}_\alpha = \{ \mathcal{V}_\alpha, \mathcal{E}_\alpha \}$ is the set of its corresponding nodes and edges.}
 Each node $v_\alpha^{(i)}$ has an associated feature vector 
 {$h_\alpha^{(il)}$}
 , and~the entire graph has an associated edge attribute tensor $a_\alpha^{(ijr)}$ describing $r$ different relationships between node $v_\alpha^{(i)}$ and its neighbors $v_\alpha^{(j)}$. Here, we can define $\mathcal{N}(i)$ as the set of neighbors of node $v_\alpha^{(i)}$ and take $r=1$, as~we only consider one edge attribute
 {. In~other words, the~edge attribute tensor $a_\alpha^{(ijr)} \to a_\alpha^{(ij)}$ becomes a matrix.}
 The edge attributes are typically formed from the features corresponding to each node and its~neighbors.

\begin{figure}[H]
  \includegraphics[width=0.42\textwidth]{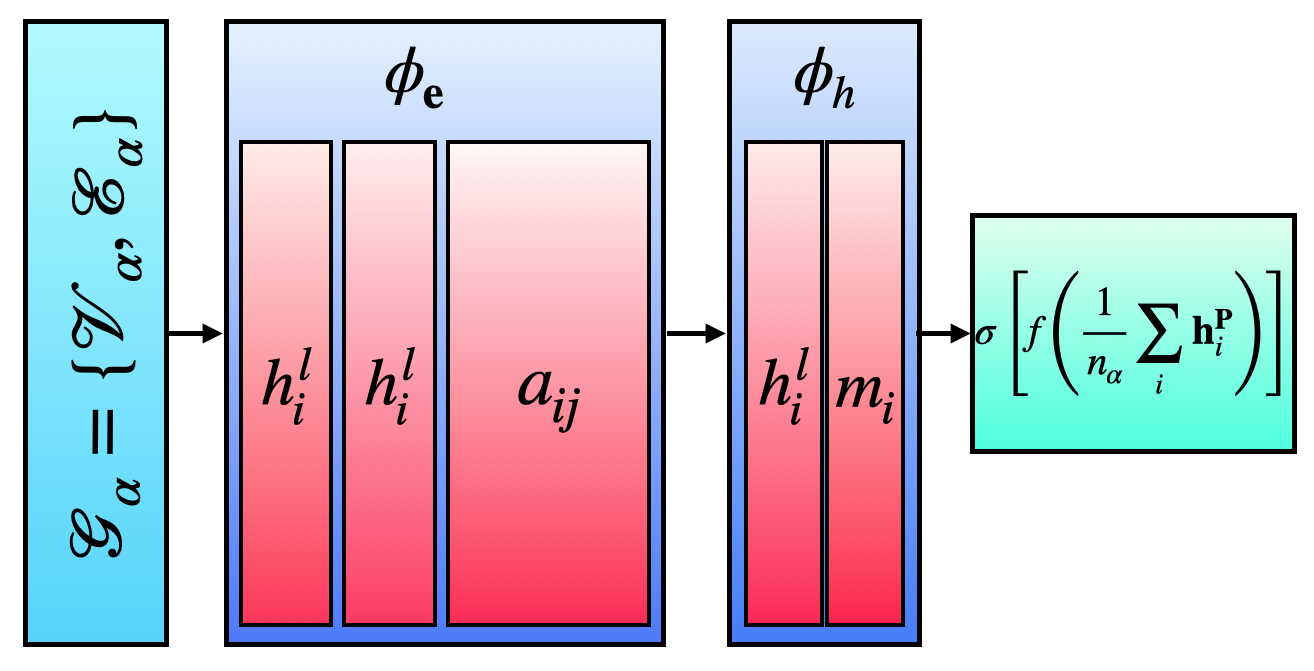} 
  \includegraphics[width=0.55\textwidth]{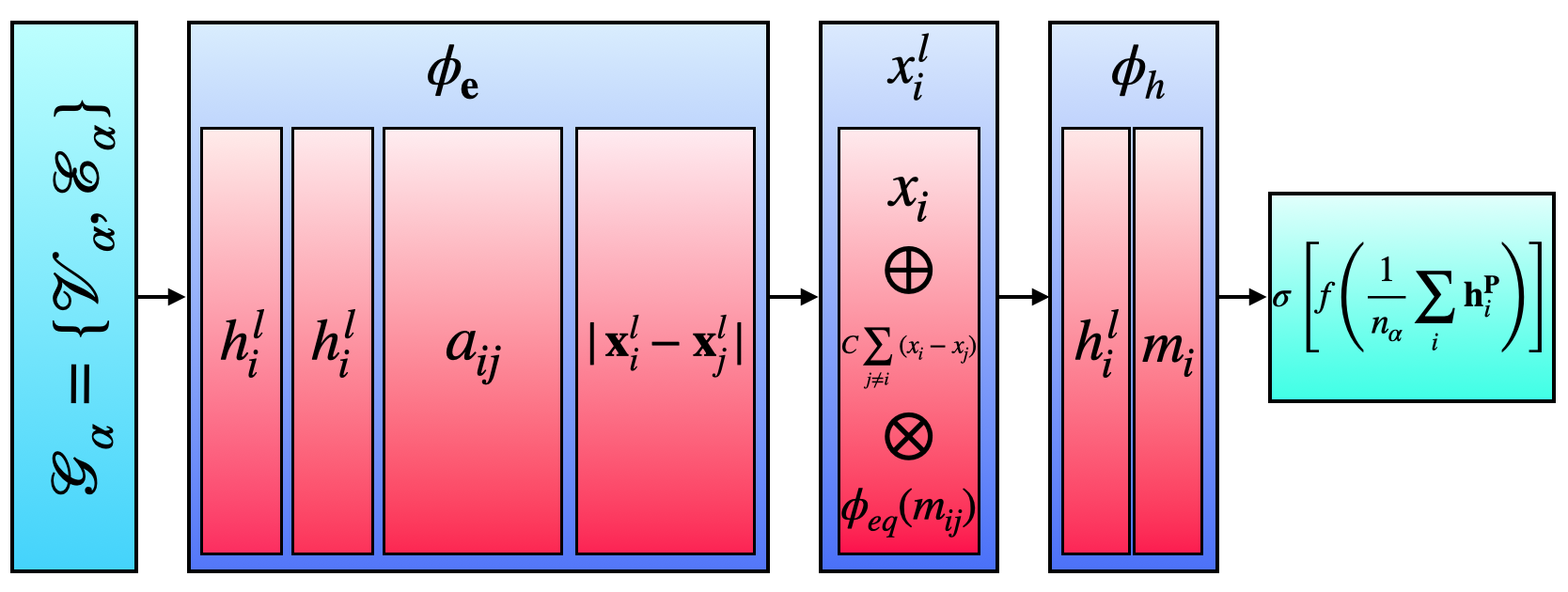}
  \caption{Graph neural network (GNN, \textbf{left}) and equivariant graph neural network (EGNN, \textbf{right}) schematic~diagrams.}
  \label{fig3:gnn_egnn}
\end{figure}

In the layer structure of a GNN, multilayer perceptions (MLPs) are used to update the node features and edge attributes before aggregating, or~mean pooling, the~updated node features for each graph to make a final prediction. To~simplify notation, we omit the graph index $\alpha$, lower the node index $i$, and~introduce a model layer index $l$. The~first MLP is the edge MLP $\phi_e$, which, at~each layer $l$, collects the features $\mathbf{h}_i^l$, its neighbors’ features $\mathbf{h}_j^l$, and~the edge attribute $a_{ij}$ corresponding to the pair. Once the new edge matrix $m_{ij}$ is formed, we sum along the neighbor dimension $j$ to create a new node feature $\mathbf{m}_i$. This extra feature is then added to the original node features $\mathbf{h}_i$ before being input into a second node updating MLP $\phi_h$ to form new node features $h_i^{l+1}$ \cite{kipf2016semi,gilmer2017neural,satorras2022en}. Therefore, a~GNN is defined {as}
\begin{align}
  \mathbf{m}_{ij} &= \phi_e(\mathbf{h}_i^l,\mathbf{h}_j^l,a_{ij}), \notag \\
   \mathbf{m}_i &= \sum_{j \in \mathcal{N}(i)} \mathbf{m}_{ij}, \\
   \mathbf{h}_i^{l+1} &= \phi_h(\mathbf{h}_i^l,\mathbf{m}_i). \notag
\end{align}
{Here}, $h_i^l \in \mathbb{R}^k$ is the $k$th-dimensional embedding of node $v_i$ at layer $l$, and $\mathbf{m}_{ij}$ is typically referred to as the message-passing function. Once the data are sent through the $P$ graph layers of the GNN, the~updated nodes $\mathbf{h}_i^{P}$ are aggregated via mean pooling for each graph to form a set of final features $\frac{1}{n_\alpha} \sum_{i=1}^{n_\alpha}\mathbf{h}_i^{P}$. These final features are sent through a fully connected neural network (NN) to output the predictions. Typically, a~fixed number of hidden features $k = N_h$ is defined for both the edge and node MLPs. The~described GNN architecture is pictorially shown in the left panel in Figure~\ref{fig3:gnn_egnn}.

\subsection{SE(2) Equivariant Graph Neural Network}

For the classical EGNN, the~approach used here was informed by the successful implementation of SE(3), or~rotational, translational, and~permutational, equivariance on dynamic systems and the QM9 molecular dataset~\cite{satorras2022en}. It should be noted that GNNs are naturally permutation equivariant, in~particular invariant, when averaging over the final node feature outputs of the graph layers~\cite{thiede2020general}. An~SE(2) EGNN takes the same form as a GNN; however, the~coordinates are equivariantly updated within each graph layer, i.e.,~$\mathbf{x}_i \to \mathbf{x}_i^l$ where $\mathbf{x}_i = (\phi_i, y_i)$ in our case. The~new form of the network {becomes} 
\begin{align}
  \mathbf{m}_{ij} &= \phi_e(\mathbf{h}_i^l,\mathbf{h}_j^l,a_{ij}, |\mathbf{x}_i^l - \mathbf{x}_j^l|) , \notag \\
   \mathbf{m}_i &= \sum_{j \in \mathcal{N}(i)} \mathbf{m}_{ij}, \\
   \mathbf{x}_i^{l+1} & = \mathbf{x}_i^l + C \sum_{j \neq i} (\mathbf{x}_i^l - \mathbf{x}_j^l) \phi_x(\mathbf{m}_{ij}) , \notag \\
   \mathbf{h}_i^{l+1} &= \phi_h(\mathbf{h}_i^l,\mathbf{m}_i) . \notag
\end{align}
{Since} the coordinates $\mathbf{x}_i^l$ of each node $v_i$ are equivariantly evolving, we also introduce a second invariant embedding $|\mathbf{x}_i^l - \mathbf{x}_j^l|$ based on the equivariant coordinates into the edge MLP $\phi_e$. The~coordinates $\mathbf{x}_i$ are updated by adding the summed difference between 
{the coordinates of}
 $\mathbf{x}_i$ and its neighbors $\mathbf{x}_j$
 {. This added term is 
 {suppressed}
  by a factor of $C$, which we take to be $C(n_\alpha) = \frac{1}{\ln(2n_\alpha)}$. The~term is further multiplied by a coordinate MLP $\phi_x,$ which takes as input the message-passing function $\mathbf{m}_{ij}$ between node $i$ and its neighbors $j$.}
  For a proof 
  {of}
  the rotational and translational equivariance of $\mathbf{x}_i^{l+1}$, see Appendix \ref{appendixA:equivariance}. The~described EGNN architecture is pictorially shown in the right panel in Figure~\ref{fig3:gnn_egnn}.

\subsection{Quantum Graph Neural~Network}

For a QGNN, the~input data, as~a collection of graphs $\{ \mathcal{G}_\alpha \}$, are the same as described above. We fix the number of qubits $n$ to be the number of nodes $n_\alpha$ in each graph. For~the quantum algorithm, we first form a normalized qubit product state from an embedding MLP $\phi_{|\psi^0 \rangle}$, which takes in the features $\mathbf{h}_i$ of each node $v_i$ and reduces each of them to a qubit state $| \phi_{|\psi^0 \rangle} (\mathbf{h}_i) \rangle \in \mathbb{C}^2$ \cite{mernyei2022equivariant}. The~initial product state describing the system then becomes $| \psi_\alpha^0 \rangle = \bigotimes_{i=1}^n | \phi_{|\psi\rangle} (\mathbf{h}_i) \rangle \in \mathbb{C}^{2^n}$, which we normalize by $\sqrt{ \langle \psi_\alpha^0 | \psi_\alpha^0 \rangle}$.

\begin{figure}[H]
  \includegraphics[width=0.34\textwidth]{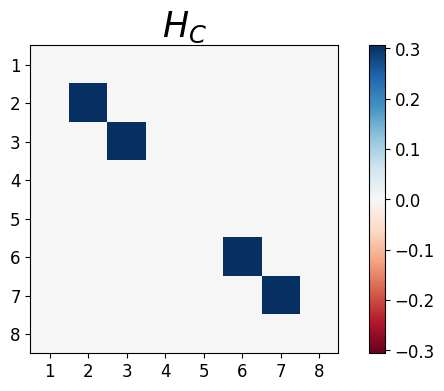}
  \includegraphics[width=0.34\textwidth]{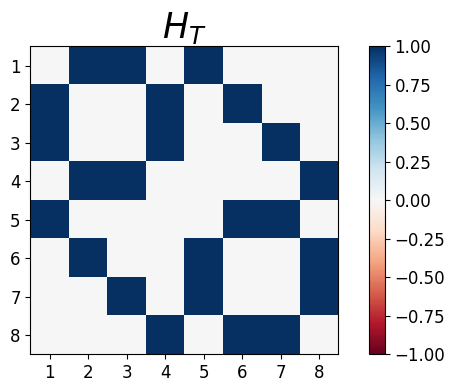}
  \caption{The $8 \times 8$ matrix representations of the coupling and transverse Hamiltonians defined in Equation~(\ref{eq:hamiltonian}).
  }
  \label{fig:Hamiltonians}
\end{figure}

{A fully parameterized Hamiltonian can then be constructed based on the adjacency matrix $a_{ij}$, or~trainable interaction weights $\mathcal{W}_{ij}$, and~node features $\mathbf{h}_i$, or~trainable feature weights $\mathcal{M}_i$ \cite{verdon2019quantum}.}
 Here, for~the coupling term of the Hamiltonian $H_C$, we utilize the edge matrix $a_{ij}$ connected to two coupled Pauli-Z operators, $\sigma_i^z$ and $\sigma_j^z$, which act on the Hilbert spaces of qubits $i$ and $j$, respectively. Since we embed the quantum state $| \psi^0 \rangle$ based on the node features $\mathbf{h}_i$, we omit the self-interaction term 
 which utilizes the chosen features applied to the Pauli-Z operator, $\sigma_i^z$, 
 {which acts}
 on the Hilbert space of qubit $i$. We also introduce a transverse term $H_T$ to each node in the form of a Pauli-X operator, $\sigma_i^x$
 {,} with a fixed or learnable constant coefficient $\mathcal{Q}_0$, which we take to be $\mathcal{Q}_0 = 1$. Note that the Hamiltonian $H$ contains $\mathcal{O}(2^n \times 2^n)$ entries due to the Kronecker products between Pauli operators. To~best express the properties of each graph, we take the Hamiltonian of the~form
\begin{align}
  H (a_{ij}) &= \underbrace{\sum_{(i,j) \in \mathcal{E}} a_{ij} \left( \frac{\hat{\mathbb{I}}_i - \sigma_i^z}{2} - \frac{\hat{\mathbb{I}}_j - \sigma_j^z}{2} \right)^2}_{H_C} + \underbrace{\sum_{i \in \mathcal{V}} \sigma_i^{x}}_{H_T},
\label{eq:hamiltonian}  
\end{align}
where the $8 \times 8$ matrix representations of $H_C$ and $H_T$ are shown in {Figure}
~\ref{fig:Hamiltonians}.
We generate the unitary form of the operator via the complex exponentiating of the Hamiltonian with real learnable coefficients $\gamma_{lq} \in \mathbb{R}^{P \times Q}$, which can be thought of as infinitesimal parameters running over $Q=2$ Hamiltonian terms and $P$ layers of the network. Therefore, the~QGNN can be defined as
\begin{subequations}
\begin{align}
    U_{ij} &= \phi_u(a_{ij}) = e^{-i \sum_{q=1}^Q \gamma_{lq} H_{q} (a_{ij}) }, \\
    |\psi^{l+1}\rangle &= \phi_{|\psi \rangle }(| \psi^l \rangle, U_{ij}) = U_\theta^l U_{ij} U_\theta^{l^\dagger} | \psi^l \rangle ,
\end{align}
\end{subequations}
where $U_\theta^l = (\vect{\theta}' - i\mathbb{I})(\vect{\theta}' + i\mathbb{I})^{-1}$ is a parameterized unitary Cayley transformation in which we force $\vect{\theta}' = \vect{\theta} + \vect{\theta}^\dagger$ to be self-adjoint, i.e.,~$\vect{\theta}' = \vect{\theta}'^\dagger$, with~$\vect{\theta} \in \mathbb{C}^{2^n \times 2^n}$ as the trainable parameters. The~QGNN evolves the quantum state $|\psi^0 \rangle$ by applying unitarily transformed ansatz Hamiltonians with $Q$ terms to the state over $P$ layers. 

\vspace{-5pt}

\begin{figure}[H]
  \includegraphics[width=0.49\textwidth]{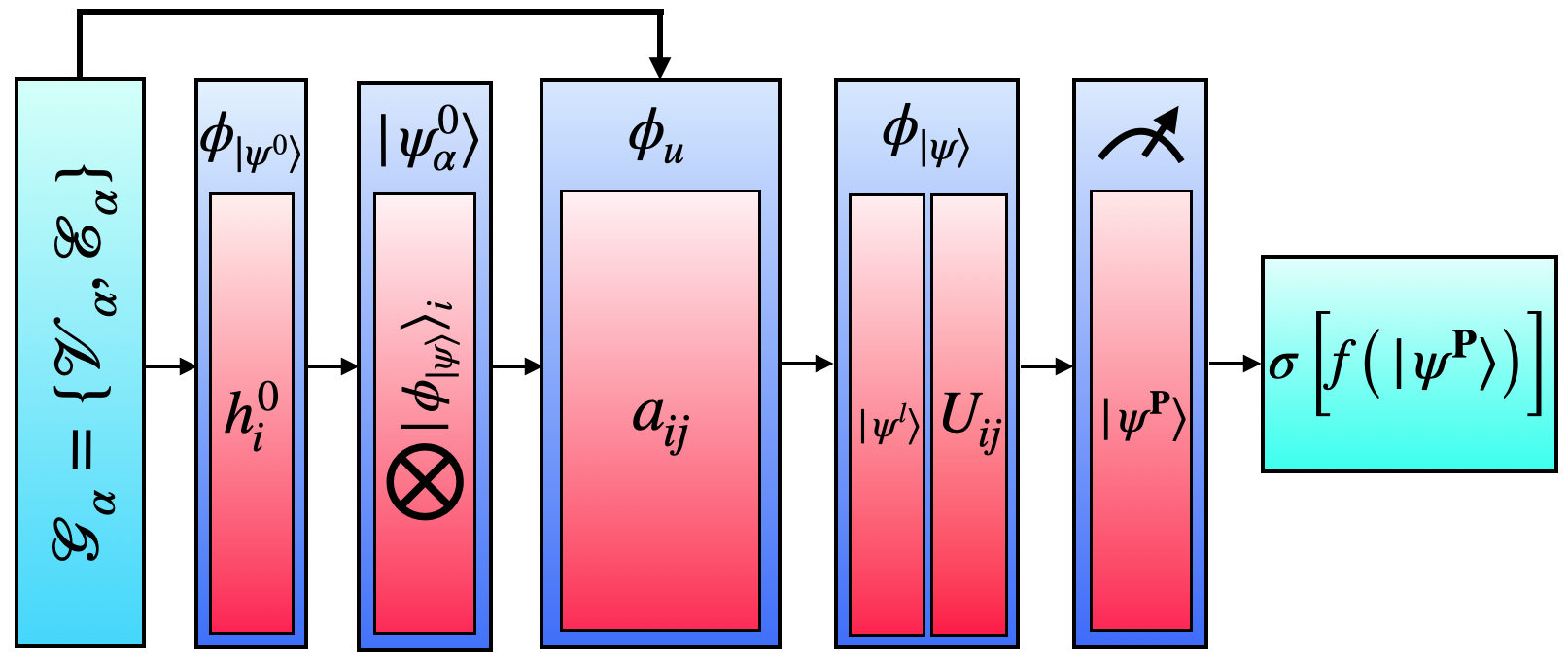}
  \includegraphics[width=0.49\textwidth]{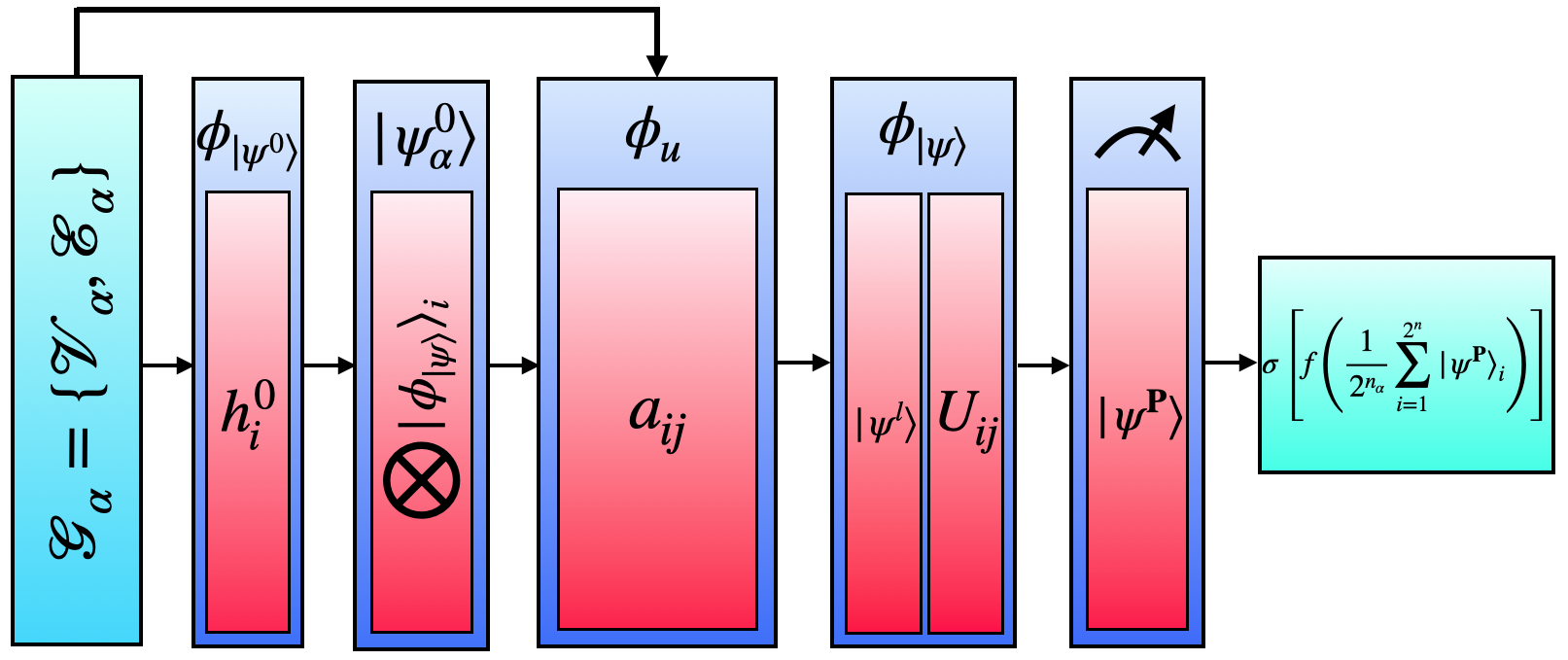}
  \caption{{Quantum} 
 graph neural network (QGNN, \textbf{left}) and equivariant quantum graph neural network (EQGNN, \textbf{right}) schematic~diagrams.}
  \label{fig:quantum_networks}
\end{figure}

The~final product state $|\psi^P \rangle \in \mathbb{C}^{2^n}$ is measured for output, which is sent to a classical fully connected NN to make a prediction. The~analogy between the quantum unitary interaction function $\phi_u$ and classical edge MLP $\phi_e$, as~well as between~the quantum unitary state update function $\phi_{|\psi\rangle}$ and classical node update function $\phi_h$, should be clear. For~a reduction in the coupling Hamiltonian $H_C$ in Equation~(\ref{eq:hamiltonian}), see Appendix \ref{appendixB:hamiltonian}. The described QGNN architecture is pictorially shown in the left panel in {Figure}~\ref{fig:quantum_networks}.

\subsection{Permutation Equivariant Quantum Graph Neural~Network}
The EQGNN takes the same form as the QGNN; however, we aggregate the final elements of the product state $\frac{1}{2^n} \sum_{k=1}^{2^n}| \psi^P_k \rangle$ via mean pooling before sending this complex value to a fully connected NN~\cite{nguyen2022theory,mernyei2022equivariant,skolik2023}. See Appendix \ref{appendix:quantum_equivariance} for a proof of the quantum product state permutation equivariance over the sum of its elements. The~resulting EQGNN architecture is shown in the right panel in {Figure}~\ref{fig:quantum_networks}.

\section{Results and~Analysis}
\begin{figure}[H]
\begin{adjustwidth}{-\extralength}{0cm}
      \centering
      \includegraphics[width=0.31\textwidth]{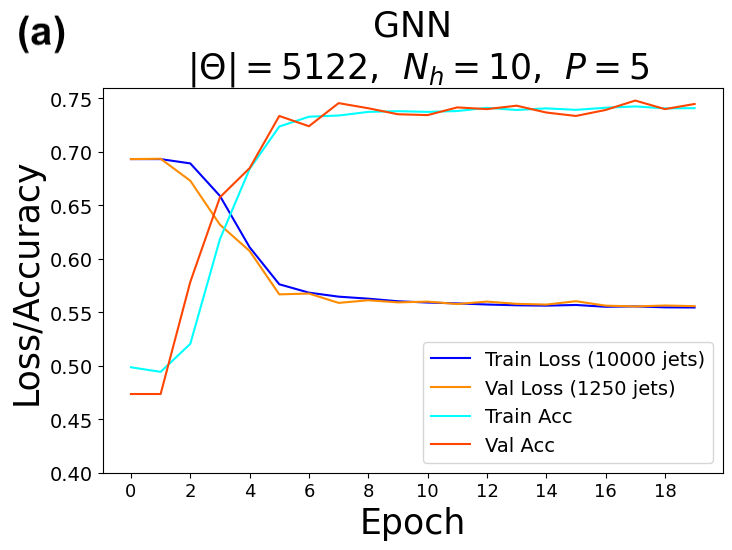}
      \includegraphics[width=0.31\textwidth]{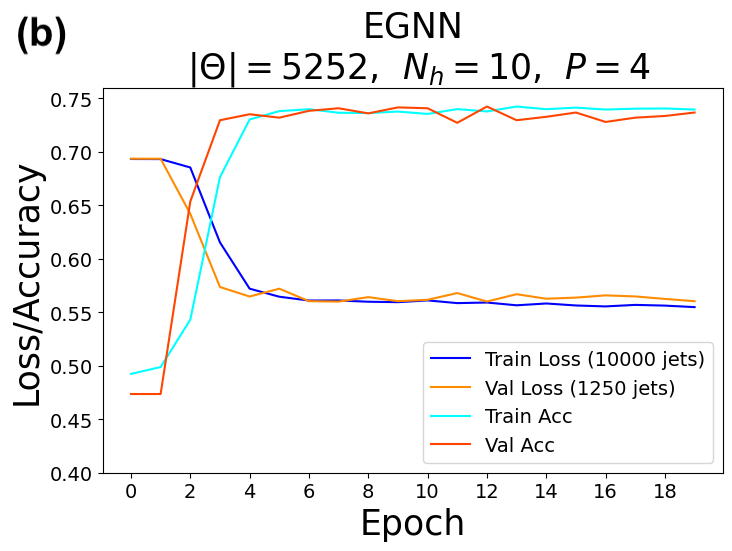}
      \includegraphics[width=0.31\textwidth]{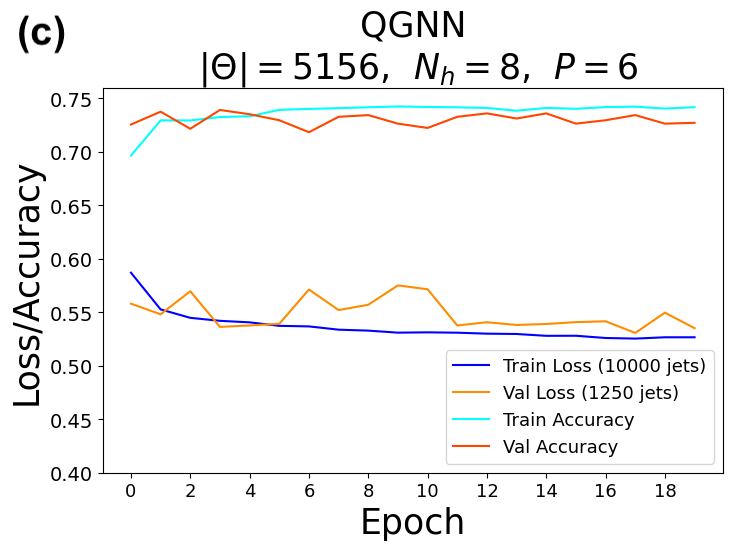} 
      \includegraphics[width=0.31\textwidth]{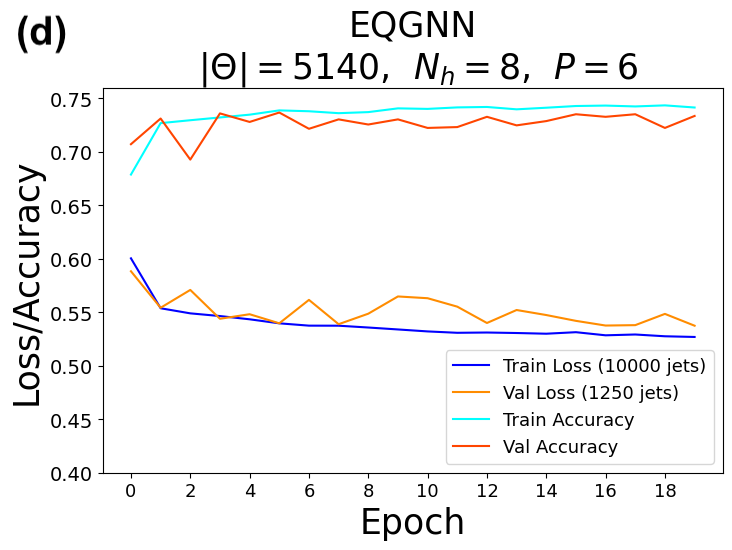}
\end{adjustwidth}
      \caption{{(\textbf{a}) GNN,} 
       (\textbf{b}) EGNN, (\textbf{c}) QGNN, and~(\textbf{d}) EQGNN training history~plots.}
      \label{figure:training}
  \end{figure}

For each model, a~range of total parameters 
{was}
 tested; however, the~overall comparison test was conducted using the largest optimal number of total parameters for each network. A~feed-forward NN was used to reduce each network's graph layered output to a binary one, followed by the softmax activation function to obtain the logits in the classical case and the norm of the complex values to obtain the logits in the quantum case. Each model trained over $20$ epochs with the Adam optimizer consisting of a learning rate of $\eta = 10^{-3}$ and a cross-entropy loss function. The~classical networks were trained with a batch size of $64$ and the quantum networks with a batch size of one due to the limited capabilities of broadcasting unitary operators in the quantum APIs. 
 {After epoch $15$, the~best model weights were saved when the validation AUC of the true positive rate (TPR) as a function of the false positive rate (FPR) was maximized.}
 The results of the training for the largest optimal total number of parameters $|\Theta | \approx 5100$ are shown in Figure~\ref{figure:training}, with more details listed in Table~\ref{table:results}.
\unskip
\begin{table}[H]
\begin{threeparttable}[b]
 \caption{{Metric} 
 comparison between the classical and quantum graph~models \tnote{1}.}
 \label{table:results}
\newcolumntype{C}{>{\centering\arraybackslash}X}
    \begin{tabularx}{\textwidth}{CCCCCCCC} 
    \toprule
    \textbf{Model} & \boldmath{$|\Theta|$} & \boldmath{$N_{h}$} & \textbf{\emph{P}}& \textbf{Train ACC} & \textbf{Val ACC} & \textbf{ Test AUC}  \\
    \midrule
    \textbf{GNN} & 5122 & 10 & 5 & $74.25 \%$ & $74.80 \%$ & \textbf{\color{magenta} $\mathbf{63.36 \%}$} \\ 
    \textbf{EGNN} & 5252 & 10 & 4 & $73.66 \%$ &$74.08 \%$ & \textbf{\color{red} $\mathbf{67.88 \%}$}  \\ 
    \textbf{QGNN} & 5156 & 8 & 6 & $74.00 \%$ &$73.28 \%$ & \textbf{\color{blue} $\mathbf{61.43 \%}$}  \\ 
    \textbf{EQGNN} & 5140 & 8 & 6 & $74.42 \%$ &$72.56 \%$ & \textbf{\color{cyan} $\mathbf{75.17 \%}$}  \\
    \bottomrule
    \end{tabularx}
\noindent{\footnotesize{$^{1}$ The pink color represents the GNN results. The red color represents the EGNN results. The blue color represents the QGNN results. The cyan color represents the EQGNN results. This representation extends to Figure~~\ref{figure:auc_theta}.}}
\end{threeparttable}
\end{table}

Recall that for each model, we varied the number of hidden features $N_h$ in the $P$ graph layers. Since we fixed the number of nodes $n_\alpha = 3$ per jet, the~hidden feature number $N_h = 2^3 = 8$ was fixed in the quantum case, and,~therefore, we also varied the parameters of the encoder $\phi_{|\psi^0 \rangle}$ and decoder NN in the quantum~algorithms.

\begin{figure}[H]
      \includegraphics[width=0.47\textwidth]{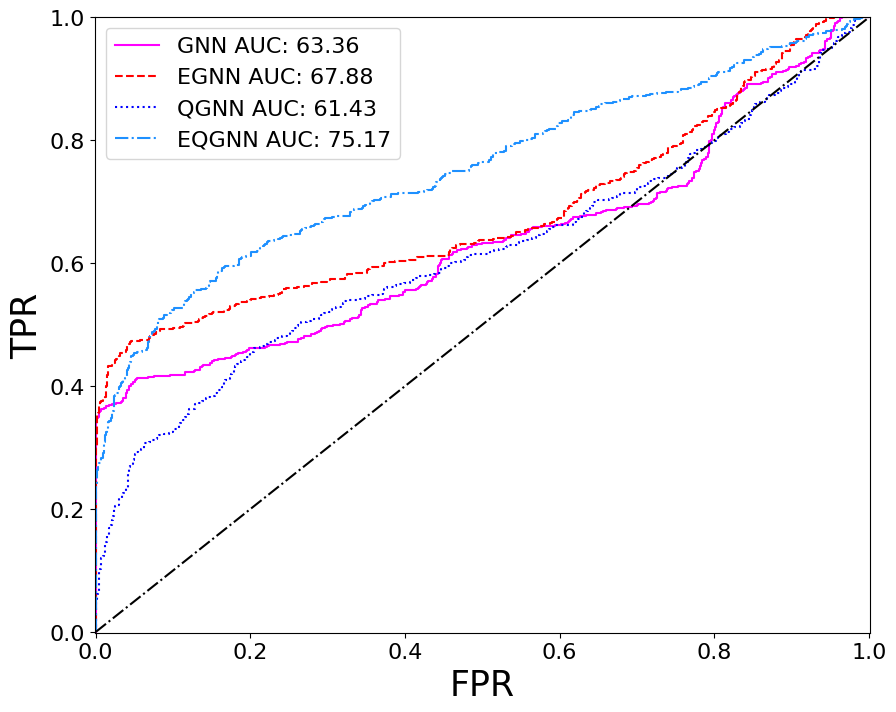}
      \includegraphics[width=0.47\textwidth]{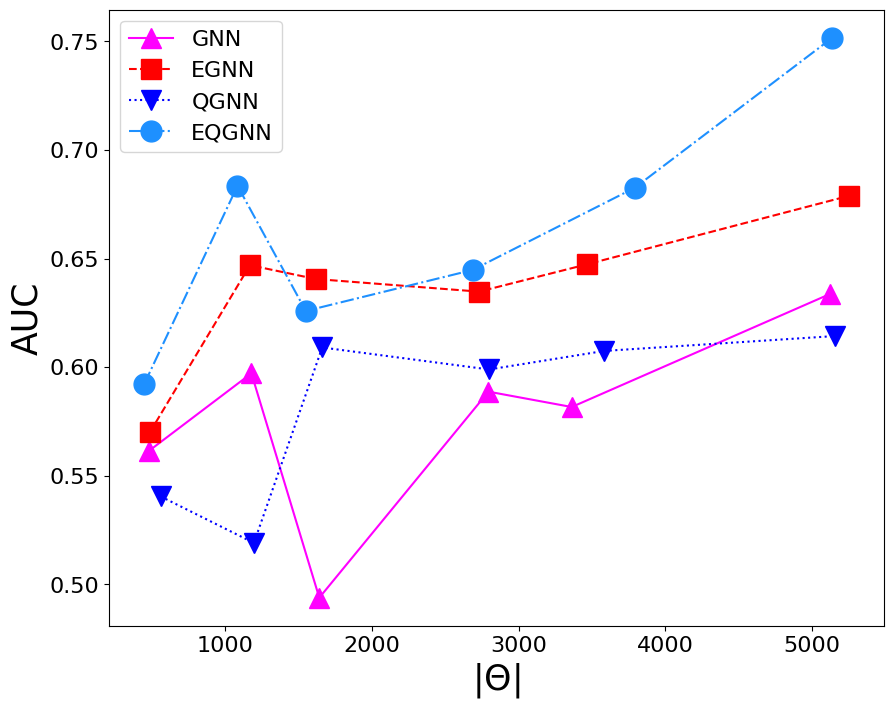}
      \caption{Model ROC curves (\textbf{left}) and AUC scores as a function of parameters (\textbf{right}).
      }
      \label{figure:auc_theta}
  \end{figure}

Based on the AUC scores, the~EGNN outperformed both the classical and quantum GNN; however, this algorithm was outperformed by EQGNN with a $7.29 \%$ increase in AUC, representing the strength of the EQGNN. Although~the GNN outperformed the QGNN in the final parameter test by $1.93 \%$, the~QGNN performed more consistently and outperformed the GNN in the mid-parameter range $|\Theta| \in (1500,4000)$. Through the variation in the number of parameters, the~AUC scores were recorded for each case, where the number of parameters taken for each point corresponded to $|\Theta| \approx \{500, 1200, 1600, 2800, 3500, 5100 \}$, as shown in the right panel in Figure~\ref{figure:auc_theta}.

\section{Conclusions}


Through several computational experiments, the~quantum GNNs seemed to exhibit enhanced classifier performance compared with their classical GNN counterparts based on the best test AUC scores produced after the training of the models while relying on a similar number of parameters, hyperparameters, and~model structures. These results seem promising for the quantum advantage over classical models. Despite this result, the~quantum algorithms took over $100$ times longer to train than the classical networks. 
{This was primarily due to the fact that we ran our quantum simulations on classical computers and not on actual quantum hardware. In~addition, we were hindered by }
 the limited capabilities in the quantum APIs, where the inability to train with broadcastable unitary operators forced the quantum models to take in a batch size of one or~train on a single graph at a~time.

The action of the equivariance in the EGNN and EQGNN could be further explored and developed. This is especially true in the quantum case where more general permutation and SU(2) equivariance have been explored~\cite{skolik2023,mernyei2022equivariant,east2023need,zheng2023sncqa}. Expanding the flexibility of the networks to an arbitrary number of nodes per graph should also offer increased robustness; however, this may continue to pose challenges in the quantum case due to the current limited flexibility of quantum software. A~variety of different forms of the networks can also be explored
{. Potential ideas for this include}
 introducing attention components and altering the structure of the quantum graph layers to achieve enhanced performance of both classical and quantum GNNs. In~particular, one can further parameterize the quantum graph layer structure to better align with the total number of parameters used in the classical structures. These avenues will be explored in future~work.


\section{Software and~Code}
\textls[-15]{PyTorch and Pennylane were the primary APIs used in the formation and testing of these algorithms. The~code corresponding to this study can be found {at} 
{\url{https://github.com/royforestano/2023\_gsoc\_ml4sci\_qmlhep\_gnn}} (accessed on 5 February 2024).}

\vspace{6pt} 



\authorcontributions{
Conceptualization, R.T.F.; 
methodology, M.C.C., G.R.D., Z.D., R.T.F., S.G., D.J., K.K., T.M., K.T.M., K.M. and E.B.U.; 
software, R.T.F.; 
validation, M.C.C., G.R.D., Z.D., R.T.F., T.M. and E.B.U.; 
formal analysis, R.T.F.; 
investigation, M.C.C., G.R.D., Z.D., R.T.F., T.M. and E.B.U.; 
resources, R.T.F., K.T.M. and K.M.; 
data curation, G.R.D., S.G. and~T.M.; 
writing---original draft preparation, R.T.F.; 
writing---review and editing, S.G., D.J., K.K., K.T.M. and K.M.; 
visualization, R.T.F.;
supervision, S.G., D.J., K.K., K.T.M. and K.M.; 
project administration, S.G., D.J., K.K., K.T.M. and K.M.; 
funding acquisition, S.G. 
All authors have read and agreed to the published version of the~manuscript.}

\funding{This study used resources of the National Energy Research Scientific Computing Center, a~DOE Office of Science User Facility supported by the Office of Science of the U.S. Department of Energy under Contract No. DE-AC02-05CH11231 using NERSC award NERSC DDR-ERCAP0025759. S.G. was supported in part by the U.S. Department of Energy (DOE) under Award No. DE-SC0012447. K.M. was supported in part by the U.S. DOE award number DE-SC0022148. K.K. was supported in part by US DOE DE-SC0024407. {Z.D.} 
 was supported in part by College of Liberal Arts and Sciences Research Fund at the University of Kansas. {Z.D.}, R.T.F., E.B.U., M.C.C., and T.M. were participants in the 2023 Google Summer of Code.}

\institutionalreview{Not applicable.}


\dataavailability{The high-energy physics (HEP) dataset {Pythia8 Quark and Gluon Jets for Energy Flow} 
\cite{Komiske2019} was used in this analysis.} 


\conflictsofinterest{The authors declare no conflicts of interest. The~funders had no role in the design of the study; in the collection, analyses, or~interpretation of data; in the writing of the manuscript; or in the decision to publish the~results.} 



\abbreviations{Abbreviations}{
The following abbreviations are used in this manuscript:\\

\noindent 
\begin{tabular}{@{}ll}
API & Application Programming Interface \\
AUC & Area Under the Curve \\
CNN & Convolutional Neural Network \\
EGNN & Equivariant Graph Neural Network\\
EQGNN & Equivariant Quantum Graph Neural Network \\
FPR & False Positive Rate \\
GAN & Generative Adversarial Network \\
GNN & Graph Neural Network \\
LHC & Large Hadron Collider\\
MDPI & Multidisciplinary Digital Publishing Institute\\
MLP & Multilayer Perceptron \\
NLP & Natural Language Processor \\
NN & Neural Network \\
QGCNN & Quantum Graph Convolutional Neural Network \\
QGNN & Quantum Graph Neural Network \\
QGRNN & Quantum Graph Recurrent Neural Network \\
TPR & True Positive Rate 
\end{tabular}
}

 \appendixtitles{yes} 
 \appendixstart
\appendix



\section{Equivariant Coordinate Update~Function}
\label{appendixA:equivariance} 
Let $T_g : X \to X$ be the set of translational and rotational group transformations with elements $g \in T_g \subset G$ that act on the vector space $X$. The~function $\varphi : X \to X$ defined by
\begin{align}
\varphi (x) = x_i + C \sum_{j\neq i}(x_i-x_j)
\end{align}
is equivariant with respect to $T_g$.
\begin{proof}
Let a general transformation $g \in T_g$ act on $X$ by $gX = RX + T$, where $R \in T_g$ denotes a general rotation, and $T \in T_g$ denotes a general translation. Then, under transformation $g$ on $X$ of function $\varphi$ defined above, we have
\begin{align*}
  \varphi (g x) &= (gx_i ) + C \sum_{j \neq i} (gx_i - gx_j ) \\
  &= (Qx_i + T) + C \sum_{j \neq i} (Qx_i +T - Qx_j - T) \\
  &= (Qx_i + T) + C \sum_{j \neq i} (Qx_i - Qx_j) \\
  &= Qx_i + C \sum_{j \neq i} Q(x_i - x_j) + T \\
  &= Q [ x_i + C \sum_{j \neq i} (x_i - x_j) ] +T \\
  &= g \varphi (x),
\end{align*}
where $\varphi (g x) = g \varphi (x)$ shows $\varphi$ transforms equivariantly under transformations $g \in T_g$ acting on $X$.
\end{proof}

\section{Coupling Hamiltonian~Simplification}
\label{appendixB:hamiltonian}
The reduction in the coupling Hamiltonian becomes
\begingroup\makeatletter\def\f@size{9.5}\check@mathfonts
\def\maketag@@@#1{\hbox{\m@th\fontsize{10}{10}\selectfont\normalfont#1}}%
\begin{align*}
  \hat{H}_C &= \frac{1}{2} \sum_{(j,k) \in \mathcal{E} } \Lambda_{jk} \left[ \left( \frac{\hat{I}_j - \sigma_j^z}{2} \right) - \left( \frac{\hat{I}_k - \sigma_k^z}{2} \right) \right]^2 \\
  &=\frac{1}{8} \sum_{(j,k) \in \mathcal{E} } \Lambda_{jk} \left[ \left( \hat{I}_j - \sigma_j^z \right) - \left( \hat{I}_k - \sigma_k^z \right) \right]^2 \\
  &= \frac{1}{8} \sum_{(j,k) \in \mathcal{E} } \Lambda_{jk} \left[ \left( \hat{I}_j - \sigma_j^z \right)^2 - \left( \hat{I}_j - \sigma_j^z \right) \left( \hat{I}_k - \sigma_k^z \right) - \left( \hat{I}_k - \sigma_k^z \right)\left( \hat{I}_j - \sigma_j^z \right) + \left( \hat{I}_k - \sigma_k^z \right)^2 \right] \\
  &= \frac{1}{8} \sum_{(j,k) \in \mathcal{E} } \Lambda_{jk} \Big[  \hat{I}_j\hat{I}_j - \hat{I}_j\sigma_j^z - \sigma_j^z\hat{I}_j + \sigma_j^z\sigma_j^z - \hat{I}_j\hat{I}_k + \hat{I}_j\sigma_k^z + \sigma_j^z\hat{I}_k - \sigma_j^z\sigma_k^z - \hat{I}_k\hat{I}_j + \hat{I}_k\sigma_j^z + \sigma_k^z\hat{I}_j \\
  & \qquad \qquad \qquad - \sigma_k^z\sigma_j^z + \hat{I}_k\hat{I}_k - \hat{I}_k\sigma_k^z - \sigma_k^z\hat{I}_k + \sigma_k^z\sigma_k^z \Big] \\
  &= \frac{1}{8} \sum_{(j,k) \in \mathcal{E} } \Lambda_{jk} \left[  \hat{I}_j\hat{I}_j - 2\hat{I}_j\sigma_j^z + \sigma_j^z\sigma_j^z - 2\hat{I}_j\hat{I}_k + 2\hat{I}_j\sigma_k^z + 2\sigma_j^z\hat{I}_k - 2\sigma_j^z\sigma_k^z + \hat{I}_k\hat{I}_k - 2\hat{I}_k\sigma_k^z + \sigma_k^z\sigma_k^z \right]\\
  &= \frac{1}{8} \sum_{(j,k) \in \mathcal{E} } \Lambda_{jk} \left[  \hat{I}_j - 2\sigma_j^z + {\sigma_j^z}^2 - 2\hat{I}_j\hat{I}_k + 2\hat{I}_j\sigma_k^z + 2\sigma_j^z\hat{I}_k - 2\sigma_j^z\sigma_k^z + \hat{I}_k - 2\sigma_k^z + {\sigma_k^z}^2 \right] \\
  &= \frac{1}{8} \sum_{(j,k) \in \mathcal{E} } \Lambda_{jk} \left[  2\hat{I}_j - 2\sigma_j^z - 2\hat{I}_j\hat{I}_k + 2\hat{I}_j\sigma_k^z + 2\sigma_j^z\hat{I}_k - 2\sigma_j^z\sigma_k^z + 2\hat{I}_k - 2\sigma_k^z \right] \\
  &= \frac{1}{4} \sum_{(j,k) \in \mathcal{E} } \Lambda_{jk} \left[  \hat{I}_j - \sigma_j^z - \hat{I}_j\hat{I}_k + \hat{I}_j\sigma_k^z + \sigma_j^z\hat{I}_k - \sigma_j^z\sigma_k^z + \hat{I}_k - \sigma_k^z \right],
\end{align*}
\endgroup
and multiplying on the left by $\hat{I}_j$ and on the right by $\hat{I}_k$ produces
\begin{align}
  \implies \hat{H}_C &= \frac{1}{4} \sum_{(j,k) \in \mathcal{E} } \Lambda_{jk} \left[ \hat{I}_j\hat{I}_k - \sigma_j^z\hat{I}_k + \hat{I}_j\sigma_k^z + \sigma_j^z\hat{I}_k - \sigma_j^z\sigma_k^z - \hat{I}_j \sigma_k^z \right] \notag \\
    &= \frac{1}{4} \sum_{(j,k) \in \mathcal{E} } \Lambda_{jk} \left[ \hat{I}_j\hat{I}_k - \sigma_j^z\sigma_k^z \right].
\end{align}

\section{Quantum Product State Permutation~Equivariance}
For $V$, a commutable vector space, the~product state $\bigotimes_{i=1}^m \mathbf{v}_i : V^n \times \cdots \times V^n \to V^{n^m}$ is permutation-equivariant with respect to the sum of its entries. We prove the $n=2$ case for all $m \in \mathbb{Z}_{>0}$.
\label{appendix:quantum_equivariance}
\begin{proof}
(By Induction) Assuming we have an $n=1$ final qubit state,
\begin{align*}
  | \psi_1 \rangle = \bigotimes_{i=1}^1 \begin{pmatrix}
    v_1^1 \\ v_1^2  \end{pmatrix} = \begin{pmatrix}
    v_1^1 \\ v_1^2  \end{pmatrix},
\end{align*}
the sum of the product state elements is trivially equivariant with respect to similar graphs. If~we have $n=2$ final qubit states, the~product state is
\begin{align*}
  | \psi_i \rangle = \bigotimes_{i=\{1,2\}} 
  \begin{pmatrix}
    v_i^1 \\ v_i^2  \end{pmatrix} = 
    \begin{pmatrix}
    v_1^1 \\ v_1^2  \end{pmatrix} \otimes 
    \begin{pmatrix}
    v_2^1 \\ v_2^2  \end{pmatrix} = 
    \begin{pmatrix}
    v_1^1v_2^1 \\ 
    v_1^2v_2^1 \\ 
    v_1^1v_2^2 \\ 
    v_1^2v_2^2 
    \end{pmatrix} ,
\end{align*}
where the sum of elements becomes
\begin{align*}
    v_1^1v_2^1 + v_1^2v_2^1 + v_1^1v_2^2 + v_1^2v_2^2 &= v_2^1 v_1^1 + v_2^1v_1^2 + v_2^2v_1^1 + v_2^2v_1^2 \\
    & = v_2^1 v_1^1 + v_2^2v_1^1 + v_2^1v_1^2 + v_2^2v_1^2 ,
\end{align*}
which is equivalent to the sum of the elements
\begin{align*}
\begin{pmatrix} v_2^1 v_1^1\\ v_2^2v_1^1 \\ v_2^1v_1^2 \\ v_2^2v_1^2 \end{pmatrix} = \begin{pmatrix}
    v_2^1 \\ v_2^2  \end{pmatrix} \otimes \begin{pmatrix}
    v_1^1 \\ v_1^2  \end{pmatrix} = \bigotimes_{i=\{2,1\}} \begin{pmatrix}
    v_i^1 \\ v_i^2  \end{pmatrix}
\end{align*}
for commutative spaces where $v_i^j \in \mathbb{C}$ and $\{1,2\},\{2,1\}$ should be regarded as ordered sets, which again shows the sum of the state elements remaining unchanged when the qubit states switch positions in the product. We now assume the statement is true for $n = N$ final qubit states and proceed to show the $N+1$ case is true. The~quantum product state over $N$ elements becomes
\begin{align}
  \bigotimes_{i=1}^N | \psi_i \rangle = \bigotimes_i \begin{pmatrix}
    v_i^1 \\ v_i^2  \end{pmatrix} &= \begin{pmatrix}
    v_1^1 \\ v_1^2
  \end{pmatrix} \otimes \begin{pmatrix}
    v_2^1 \\ v_2^2
  \end{pmatrix} \otimes \cdots \otimes \begin{pmatrix}
    v_N^1 \\ v_N^2
  \end{pmatrix} ,
\end{align}
which we assume to be permutation-equivariant over the sum of its elements. We can rewrite the form of this state as
\begin{align}
  \bigotimes_{i=1}^N | \psi_i \rangle = \begin{pmatrix}
    A_1 \\ A_2 \\ \vdots \\ A_{2^N}
  \end{pmatrix} = A_j,
\end{align}
where $A_j$ defines the $2^N$ terms in the final product state. Replacing the $i+1$th entry of the Kronecker product above with a new $N+1$th state, we have
\begin{align*}
  \bigotimes_{i=1}^{N+1} | \psi_i \rangle = 
  \bigotimes_i \begin{pmatrix}
    v_i^1 \\ v_i^2  \end{pmatrix} &= 
  \underbrace{\begin{pmatrix}
    v_1^1 \\ v_1^2
  \end{pmatrix} \otimes \begin{pmatrix}
    v_2^1 \\ v_2^2
  \end{pmatrix} \otimes \cdots \otimes 
  \begin{pmatrix}
    v_{i}^1 \\ v_{i}^2
  \end{pmatrix} \otimes 
  \begin{pmatrix}
    v_{N+1}^1 \\ v_{N+1}^2
  \end{pmatrix} \otimes \begin{pmatrix}
    v_{i+1}^1 \\ v_{i+1}^2
  \end{pmatrix} \otimes \cdots \otimes \begin{pmatrix}
    v_N^1 \\ v_N^2
  \end{pmatrix}}_{N+1 \text{ terms}} .
\end{align*}
When this occurs, this new state consisting of $2^{N+1}$ elements with the $N+1$ state in the $i+1$th entry of the product can be written in terms of the old state with groupings of the new elements in $2^{N+1-i}$ batches of $2^i$ elements, i.e.,
\begin{align} 
\label{eq:new_prod_state}
  \bigotimes_{i=1}^{N+1} | \psi_i \rangle &= 
  \begin{pmatrix}
    B_1 = A_1 v_{N+1}^1 \\
    B_2 = A_2 v_{N+1}^1 \\
    \vdots \\
    B_{2^i} = A_{2^i} v_{N+1}^1\\
    B_{2^i+1} = A_1 v_{N+1}^2 \\
    \vdots \\
    B_{2^{i+1}} = A_{2^i} v_{N+1}^2 \\
    \vdots \\
    B_{2^{N+1} - 2^{i+1} +1 } = A_{2^{N} - 2^i + 1 } v_{N+1}^1 \\
    \vdots \\
    B_{2^{N+1} - 2^i} = A_{ 2^{N} } v_{N+1}^1\\
    B_{2^{N+1} - 2^{i} +1} = A_{2^{N} - 2^i + 1 } v_{N+1}^2 \\
    \vdots \\
    B_{2^{N+1}} = A_{2^N} v_{N+1}^2 \\
  \end{pmatrix},
\end{align}
which, when summed, becomes
\begin{align}
  \sum_{k=1}^{2^{N+1}} B_k &= \sum_{j=1}^{2^{N}} A_j v_{N+1}^1 + \sum_{j=1}^{2^{N}} A_j v_{N+1}^2 \notag\\
  &= (v_{N+1}^1 + v_{N+1}^2) \sum_{j=1}^{2^{N}} A_j . \notag
\end{align}
However, the~$i+1$th entry is arbitrary, and,~due to the summation permutation equivariance of the initial state $\bigotimes_{i=1}^N | \psi_i \rangle$, the~sum $\sum_{j=1}^{2^{N}} A_j $ is equivariant, in~fact invariant, under~all reorderings of the elements $|\psi_i \rangle$ in the product $\bigotimes_{i=1}^N | \psi_i \rangle$. Therefore, we conclude $\bigotimes_{i=1}^{N+1} | \psi_i \rangle$ is permutation-equivariant with respect to the sum of its elements.
\end{proof}

To show a simple illustration of why \eqref{eq:new_prod_state} is true, let us take two initial states and see what happens when we insert a new state between them, i.e.,~in the $2$nd entry in the product. This should lead to $2^{2+1-1} = 2^2 = 4$ groupings of $2^{1} = 2$ elements. To~begin, we~have
\begin{align*}
  \begin{pmatrix}
    v_1^1 \\ v_1^2 
  \end{pmatrix} \otimes \begin{pmatrix}
    v_2^1 \\ v_2^2 
  \end{pmatrix} = 
  \begin{pmatrix} 
    v_1^1v_2^1 \\ 
    v_1^2v_2^1 \\ 
    v_1^1v_2^2 \\ 
    v_1^2v_2^2
  \end{pmatrix}
  = 
  \begin{pmatrix} 
    A_1 \\ 
    A_2 \\ 
    A_3 \\ 
    A_4
  \end{pmatrix},
\end{align*}
and when we insert the new third state in the $1$st entry of the product above, we have
\begin{align*}
  \begin{pmatrix}
    v_1^1 \\ v_1^2 
  \end{pmatrix} \otimes \begin{pmatrix}
    v_3^1 \\ v_3^2 
  \end{pmatrix} \otimes \begin{pmatrix}
    v_2^1 \\ v_2^2 
  \end{pmatrix} = 
  \begin{pmatrix} 
    v_1^1 v_3^1 v_2^1 \\ 
    v_1^2 v_3^1 v_2^1\\ 
    v_1^1 v_3^2 v_2^1 \\ 
    v_1^2 v_3^2 v_2^1 \\
    v_1^1 v_3^1 v_2^2 \\ 
    v_1^2 v_3^1 v_2^2 \\ 
    v_1^1 v_3^2 v_2^2 \\ 
    v_1^2 v_3^2 v_2^2
  \end{pmatrix}
  = 
  \begin{pmatrix} 
    A_1v_3^1 \\ 
    A_2v_3^1 \\
    A_1v_3^2 \\ 
    A_2v_3^2 \\
    A_3v_3^1 \\ 
    A_4v_3^1 \\ 
    A_3v_3^2 \\ 
    A_4v_3^2 
  \end{pmatrix},
\end{align*}
which sums to 
\begin{align*}
  (A_1+A_2+A_3+A_4)v_3^1 + (A_1+A_2+A_3+A_4)v_3^2 = (v_3^1 + v_3^2)\sum_{i=1}^{2^{N} = 2^2 = 4} A_j .
\end{align*}

\begin{adjustwidth}{-\extralength}{0cm}

\reftitle{References}

\PublishersNote{}
\end{adjustwidth}
\end{document}